# Plastic avalanches in metal-organic framework crystals

Jin Zhang[∥]*, Jin Ke[∥], Bing Wang, Ximing Chen

School of Science, Harbin Institute of Technology, Shenzhen 518055, PR China

[∥]Contributed equally.
*Corresponding author. E-mail address: jinzhang@hit.edu.cn (J. Zhang).

**Abstract:** The compressive properties of metal-organic framework (MOF) crystals are not only crucial for their densification but also key in determining their performance in many applications. We herein investigated the mechanical responses of a classic crystalline MOF, HKUST-1 by using *in situ* compression tests. A serrated flow accompanied by the unique strain avalanches was found in individual and contacting crystals before their final flattening or fracture with splitting cracks. The plastic flow with serrations is ascribed to the dynamic phase mixing due to the progressive and irreversible local phase transition in HKUST-1 crystals, as revealed by molecular dynamics and finite element simulations. Such pressure-induced phase coexistence in HKUST-1 crystals also induces a significant loading-history dependence of their Young's modulus. The observation of plastic avalanches in HKUST-1 crystals here not only expands our current understanding of the plasticity of MOF crystals but also unveils a novel mechanism for the avalanches and plastic flow in crystal plasticity.

## 1. Introduction

Metal-organic frameworks (MOFs), a new type of porous organic-inorganic composite crystals, have many competitive advantages when compared to traditional porous materials. For instance, the extremely high porosity, large specific surface area, adjustable pore size and vast structural diversity of MOFs make them have broad application prospects in many fields including multiphase catalysis, gas storage and separation, chemical sensing, drug transportation, and environmental purification.[1-4] Although MOFs have considerable potential, they have not yet been utilized in industrial applications, because MOFs are usually acquired in the form of loose powders, which has limited their availability at the commercial scale.[5] Thus, MOFs are needed to be pressed into solid tablets with high density for industrial applications such as heterogeneous catalytic reaction and fluid separation.[6,7] In addition, intrinsic responses of MOFs to the mechanical pressure are inevitable in the applications of gas sorption,[8,9] shock absorption,[10,11] mechanocalorics,[12] to name a few. Under these circumstances, understanding the mechanical responses of MOFs under compression is crucial for the shaping process and also the engineering applications of MOFs.[5,6]

To date, mechanical behaviours of MOFs have been investigated by various experiments and simulations. As a popular quantum-mechanical method having the ability to describe solid-state materials, density functional theory (DFT) calculations were widely used to evaluate the mechanical properties especially elastic constants of various MOFs.[14] The elastic anisotropy of some typical MOFs including MOF-5, ZIF-8, HKUST-1, and MIL-53 has been theoretically confirmed by DFT studies.[15-18] Nevertheless, DFT calculations are computationally expensive, which are thus out of the reach for the MOFs with large pores or complex structures. Moreover, it is difficult to take into account the effects of crystal size and finite temperature in the DFT approach. To overcome the limitations of DFT calculations, molecular dynamics (MD) simulations were also employed to investigate the mechanical properties and deformations of various MOF crystals. The significant impact of various factors such as temperature and defects on mechanical properties of MOFs has been observed in MD simulations,[19-21] while responses of MOFs under various mechanical stimuli including compression, shearing, and shock also have been investigated.[22-26] Although MD can simulate the MOF crystals with a larger size, the size of the largest MOF models that can be effectively handled by MD simulations is still much smaller than that in reality.[14,23] This size discrepancy may cause some quantitative errors due to the so-called size effect.[27,28] Thus, experimental measurements are more direct and reliable in examining the mechanical properties of MOFs, among which the nanoindentation test based on whether instrumented indentation or atomic force microscopy is the most popular method.[29,30] Although various mechanical properties including the Young's modulus, hardness and fracture toughness can be captured through nanoindentation, only the local rather than overall mechanical properties of MOFs can be achieved, since a sharp tip is used in the nanoindentation to probe the mechanical properties nearby.[29] The *in situ* compression technique inside the scanning electron microscope (SEM) or transmission electron microscope can be an alternative for characterising the overall mechanical responses of powders like MOF crystals,[31,32] but the *in situ* compression study of MOFs is still in its infancy. To our best knowledge, only a few MOFs (ZIF-8[33] and UiO-type MOFs[33,35]) have been investigated by *in situ* compression tests. In addition, only individual MOF particles were considered in existing *in situ* compression studies, while in many practical applications such as tabletization, glassification and shock absorption the compression is directly applied to packed MOF crystals rather than individual crystals.[5,11] The mechanical response of the packed (or contacting) MOF crystals under compression, however, is still unknown.

Taking the representative MOF, HKUST-1 as a typical example, we investigated here the mechanical behaviours of individual and contacting HKUST-1 crystals under compression by using the *in situ* compression tests, which exhibited a unique serrated flow phenomenon after yielding. The mechanism of the plastic flow with serrations was also revealed through a series of simulations based on MD and finite element method (FEM).

**2. Results and discussion**

2.1 Response of individual MOF crystals to compression

As shown by the optical microscope image in Figure S1, most of the synthesised HKUST-1 crystals exhibit the well-defined octahedral morphology, which makes their facets possess the shape of equilateral triangle (Figure 1a) and orientate along the {111} direction (Figure 1b). The powder X-ray diffraction is in agreement with the simulated pattern (Figure S2), which confirms that the synthesised HKUST-1 are crystalline. Most HKUST-1 crystals obtained here have the edge length in the range of 4-20 μm, as determined from the confocal microscope. We first compressed individual HKUST-1 particles. To study their overall mechanical response, HKUST-1 particles with edge lengths of ~6 μm and ~10 μm were considered here, which are, respectively, comparable to and smaller than the diameter of flat punch tip (Figure 1c). In *in situ* compression tests, only the load-displacement data can be directly acquired. To better evaluate the mechanical properties of HKUST-1 crystals, we estimated their stress and strain from the obtained load-displacement data after approximating the octahedral particle as a column circumventing it as shown in Figure 1d. Definitions of stress and strain are detailed in the Supporting Information.

As for the crystal with an edge length of 10 μm, at the very beginning of loading or the just contact between the indenter tip and the particle surface, a very short nonlinear relationship is observed (Figure 1e), which is attributed to the finite radius of the indenter tip though it is ideally assumed to be flat. This can be confirmed by the good agreement between the nonlinear curve and the Hertzian contact theory (Figure S3) or the concave surface of the particle after the release of the first round of loading (Figure S4). The stress is subsequently found to linearly increase as the strain grows, indicating the elastic deformation of the HKUST-1 crystal. After the stress is larger than 222 MPa, the strain turns to dramatically increase as the stress slightly grows, deviating from the initial linear stress-strain relationship, which indicates the onset of plastic flow at the critical stress of 222 MPa. The critical stress at the elastic-to-plastic transition is termed the yield strength. To figure out whether the electron beam irradiation can affect the mechanical properties of HKUST-1 crystals, we also conducted a similar compression test to the crystal with a similar edge length of 10 μm by turning off the electron beam in the compression process. As shown in Figure S5, the yield strength now is 237 MPa, very close to the result measured with the electron beam on. The trivial effect of the electron beam irradiation in the present study is ascribed to its short dosing time and small accelerating voltage. After the yield of HKUST-1 crystals, the plastic flow or the large-strain deformation continues until the strain is around 0.37. This long plastic flow process indicates the potential mechanical energy absorption application of HKUST-1 crystals like some other MOFs such as MOF-5, UiO-type MOFs and ZIF-8.[10,11,26] Specifically, the energy of absorption defined as the integral of the corresponding stress-strain response in the plastic flow is around 82 MJ/m$^3$ for the {111}-oriented HKUST-1 crystals considered now, as calculated from Figure 1e. In the HKUST-1 crystal with an edge length of 6 μm, a similar stress-strain relationship with a long plastic flow process was also observed except for a larger yield strength with a magnitude of 257 MPa observed now probably due to the size effect.[27,28,35] Moreover, the yield strength predicted from the present HKUST-1 crystal is close to but smaller than the value of 280 MPa obtained from the previous *in situ* compression experiments of HKUST-1 micropillars whose diameter (5 μm) is comparable to the edge length of the present HKUST-1 crystal.[36] The larger yield strength observed in HKUST-1 micropillars probably is similarly due to the size effect and/or was artificially induced during the micropillar fabrication because the ion irradiation usually can stiffen materials.[37,38]

As the strain further grows at the end of plastic flow, the stress in the HKUST-1 crystal with an edge length of 10 μm is found to increase with a much larger rate, indicating the occurrence of work hardening due to the condensation of MOFs under pressure, which will be discussed latter. However, after the strain reaches 0.52, an abrupt drop of stress is observed, which dedicates the fracture of the HKUST-1 crystal with an inclined shear band or crack as shown in Figure 1f. Actually, before the final fracture, an obvious slip was observed in the HKUST-1 crystal at the strain of 0.44 (Figure 1f or Movie 1), though the stress-strain curve exhibits no significant change at this strain. However, no obvious splitting cracks were observed in a smaller HKUST-1 crystal with the edge length of 6 μm, which exhibited a squeezed flattening at the end of plastic flow (Figure 1g or Movie 2). A similar 'flatten' failure mode was also observed in the previous *in situ* compression tests of some other MOF crystals such as UiO-66 and ZIF-8.[33,35] The different fracture behaviours observed in different MOF crystals are probably attributed to their different sizes. As for the MOF crystal assumed as a column, there exists a critical diameter $d_{cr}$, which has the following expression:[39]

$$d_{cr} = \left( \frac{1}{k_S \beta} \cdot \frac{K_{IC}}{\sigma_Y} \right)^2, \tag{1}$$

where $k_S$ is a constant determined by the angle between the slip plane normal and compression axis, $\beta$ is a geometrically dependent constant, $K_{IC}$ is the fracture toughness and $\sigma_Y$ is the yield strength. The splitting can occur in the column only when its diameter is larger than $d_{cr}$. Based on the values of $K_{IC}$ and $\sigma_Y$ measured in the previous experiment,[36] $d_{cr}$ of the present {111}-oriented HKUST-1 crystals was estimated to be several micrometres to tens of micrometres. The splitting cracking is thus expected to occur in larger MOF microcrystals but will disappear from their counterparts possessing a much smaller edge length.[33,35]

2.2 Plastic avalanches and plastic flow

As shown in Figures 1e and S5, stress-strain curves after the yield point are characterized by the serrated yielding. This serrated flow behaviour is termed plastic avalanche.[40,41] The similar phenomenon was widely observed in the compression tests of metallic nanopillars, which is also called the Portevin-Le Chatelier effect.[42,43] The plastic avalanche or serrated flow phenomenon was usually ascribed to the stick-slip nature of the sliding of shear bands in metals.[40-43] However, as shown in Figures S6, 1f and 1g (or Movies 1-3), no shear bands are observed in the plastic flow process of the present HKUST-1 crystals, which indicates that there must exist some other mechanisms for the plastic avalanches in HKUST-1 crystals.

To explain the plastic avalanches or the serrated flow observed in the crystalline HKUST-1, the compression of the {111}-oriented HKUST-1 crystals was also simulated by MD simulations (Figure 2a). The load-displacement response during the whole compression simulation process is shown in Figure 2b. At the beginning, the load is found to linearly increase as the displacement grows, indicating the elastic deformation of the particle. However, when the displacement grows to 1.64 nm, an abrupt drop is observed in the load, which is associated with the collapse of the nanopores near the topmost surface (Figure 2c). After that, the load is found to increase again with growing displacement for a very short period. The change in the present load-displacement curve due to the collapse of local nanopores is exactly similar to the plastic avalanche observed in the above experiments. The plastic avalanche repeats for several periods, accompanied by the collapse of more local nanopores successively occurring from the surface to the inner of the particle. Such a sequence of plastic avalanches due to the collapse of local nanopores finally results in the plastic flow as shown in Figure 2b. After the end of the plastic flow at the displacement of 10.67 nm, the load is found to monotonically increase as the displacement grows without any serrations, which signifies the complete condensation of HKUST-1 due to the collapse of almost all nanopores in the entire HKUST-1 structure (Figure 2c).

The collapse of local nanopores in the serrated flow process actually corresponds to the irreversible transition of the local HKUST-1 material from the parent open pore (op) phase to the new closed pore (cp) phase, which can be verified by the different radial distribution functions of the HKUST-1 structures extracted before and after the plastic flow (Figure S7). Although the phase transition due to compression was also observed in our recent MD simulations of bulk HKUST-1,[21] it was found to occur cooperatively throughout the whole structure without any phase coexistence process. We attribute this difference to the different stress distributions in the different HKUST-1 structures considered. In the bulk HKUST-1 under compression, the stress uniformly distributes, causing the simultaneous collapse of all nanopores.[23] This also explains the disappearance of serrations from the plastic flow of HKUST-1 micropillars as shown in the recent study,[36] since the stress can similarly distribute uniformly in the column-like micropillar structures. However, the distribution of stress in the realistic HKUST-1 particles with octahedral configurations is usually nonuniform, which results in the progressive collapse of nanopores and thus the phase coexistence in HKUST-1 particles due to the quenching of local pressure.[23]

2.3 Loading-history dependence of the Young's modulus

According to the above discussion, there coexist both the parent op phase and transformed cp phase in the HKUST-1 crystals after yielding. Thus, it is of interest to investigate the effect of phase coexistence on the mechanical properties of HKUST-1 crystals. By calculating the slope of the stress-strain curve in the unloading process (Figure 3a), we estimated the Young's modulus of HKUST-1 crystals with an edge length of ~10 μm. To exclude the effect of ion irradiation, the electron beam irradiation was turned off during the compression tests. In Figure 3b, we illustrate the Young's moduli of HKUST-1 crystals previously compressed with different historical maximum strains. As shown in Figure 3b, when the historical maximum strain is smaller than 0.05, the Young's modulus of HKUST-1 crystals exhibiting almost pure op structure is around 5.52 GPa, almost independent of the historical maximum strain. The Young's modulus measured here agrees well with the value of 5.42 GPa predicted from MD simulations on the intrinsic op structure of HKUST-1.[21] After the yield occurs or the historical maximum strain becomes larger than 0.05, the Young's modulus of HKUST-1 crystals with the mixed phase is found to grow as the historical maximum strain increases, which ultimately increases to 11.94 GPa at the historical maximum strain around 0.49. However, as the historical maximum strain increases after 0.49, no significant changes are observed in the Young's modulus, because HKUST-1 crystals now turn to exhibit the pure cp structure due to the completeness of phase transition.

As a widely used experimental technique in evaluating the Young's modulus of MOFs,[30] instrumented indentation tests (IITs) were also employed here to directly measure the Young's modulus of pristine HKUST-1 crystals in absence of preloads (Figure S8). The obtained results together with previously reported results from similar IITs[36] are also illustrated in Figure 3b. It is found that the Young's modulus extracted from IITs is much larger than the intrinsic Young's modulus of HKUST-1 crystals with the parent cp phase but actually locates in the value range of the HKUST-1 crystals with the mixed phase. This finding denotes that the widely used IITs are difficult to capture the intrinsic mechanical properties of HKUST-1 crystals or more flexible MOFs, because the phase transition can inevitably occur in the flexible MOF materials near the indenter tip due to the high pressure generated nearby, making the tested region probably exhibit the mixed phase.[44]

Theoretically, the MOF with the mixed phase after yielding can be modelled as a composite material comprised of the op component with the Young's modulus of $E_{op}$ and the cp component with the Young's modulus of $E_{cp}$ (Figure 3c). Thus, based on the rule of mixtures for composite materials,[45] the equivalent Young's modulus $E$ of HKUST-1 crystals with a historical maximum strain of $\varepsilon_m$ can be expressed as (see the Supporting Information):

$$E(\varepsilon_m) = \begin{cases} E_{op}, & \varepsilon_m < \varepsilon_s \\ \dfrac{\varepsilon_f - \varepsilon_m}{\varepsilon_f - \varepsilon_s} E_{op} + \dfrac{\varepsilon_m - \varepsilon_s}{\varepsilon_f - \varepsilon_s} E_{cp}, & \varepsilon_s < \varepsilon_m < \varepsilon_f \\ E_{cp}, & \varepsilon_m > \varepsilon_f \end{cases} \quad (2)$$

where $\varepsilon_s$ and $\varepsilon_f$ correspond to the strains of the start and end of the plastic flow, respectively. As shown in Figure 3b, the present composite model can agree well with the experimental results.

Since the yielded HKUST-1 crystals usually exhibit the phase coexistence, it is thus of great interest to identify locations of the parent op phase and transformed cp phase in realistic particles. As an initial attempt, we used FEM together with the elastic-perfectly plastic model to simulate the mechanical responses of an HKUST-1 particle with the dimension comparable to that considered in experiments. The load-displacement output of FEM simulations is found to be overlaid on experimental load-displacement curves (Figure S9), despite its failure in capturing the avalanche phenomenon of plasticity. Figure 3d shows that the deformation initiates majorly near the topmost and bottommost surfaces and propagates inwards the particle progressively. This result together with the corresponding stress distribution shown in Figure S10 indicates that the op-to-cp phase transition will start from the surfaces of particles and spread into the inside during the compression process, which can be further confirmed by the large residual strains or the cp-phased regions left near the surfaces of particles after the complete release of compressive loading. The evolution and distribution of the cp phase in HKUST-1 crystals predicted now by FEM simulations are identical to the results obtained from the above MD simulations though the size of the crystal considered in MD simulations is much smaller than that in reality.

2.4 Serrated flow in the plasticity of contacting MOF crystals

In many practical applications, packed rather than individual MOF crystals are subjected to the compression. Thus, to verify whether the major conclusion extracted above from individual HKUST-1 particles can be extended to the packed particles, *in situ* compression tests were also performed to the contacting HKUST-1 crystals. In doing this, as shown in Figure 4a, the indenter tip was first in contact with an HKUST-1 particle. Due to the larger interaction between the HKUST-1 particle and the diamond indenter tip, the HKUST-1 particle detached from the substrate and attached to the indenter tip. After that, the indenter tip adhered with an HKUST-1 particle was moved exactly above another HKUST-1 particle to make these two particles align with each other. Finally, the indenter tip was gradually moved downwards to realise the compression of the contacting HKUST-1 particles.

Two cases that the contacting particles have comparable and different sizes were considered in the present study (inset of Figure 4a). As shown in Figure 4b and 4c, the load-displacement relationships of these two cases are analogous to each other, both of which exhibit a serrated flow behaviour similar to the individual crystals. Nevertheless, HKUST-1 particles in these two cases exhibit different deformation behaviours. As shown in Figure 4d (or Movie 4), if the contacting HKUST-1 particles have comparable sizes, their deformations are synchronous, since these two particles have the comparable stiffness (inset of Figure 4b). As a result, the serrated flow now is induced by the op-to-cp phase transition simultaneously occurring in both particles. However, if the contacting HKUST-1 particles have different sizes, the phase transition majorly occurs in the smaller particle first (Figure 4e or Movie 5), since it possesses a smaller equivalent stiffness (inset of Figure 4c). After the almost completeness of the phase transition in the smaller particle, the deformation of the larger particle becomes dominant. In other words, the serrated flow in the contacting HKUST-1 particles with different sizes is caused by the op-to-cp phase transition successively occurring from the smaller particles to larger ones.

## 3. Conclusions

The mechanical responses of individual and contacting HKUST-1 crystals under compression were investigated by *in situ* compression tests. A plastic flow accompanied by unique strain avalanches was observed in the crystals before their final flattening or fracture with splitting cracks. Through MD and FEM simulations, it is found that the plastic avalanches in HKUST-1 crystals are caused by the dynamic phase mixing originating from the progressive and irreversible closure of their nanopores. The sequence of such strain avalanches broadly distributed in particles results in the overall plastic flow of HKUST-1 crystals. In addition, due to the pressure-induced phase coexistence in the yielded HKUST-1 crystals, the Young's modulus of HKUST-1 crystals is found to strongly depend on the loading history, which increases as the historical maximum strain grows. The present work not only significantly expands current knowledge of mechanical properties especially the plasticity of MOF crystals, but also discloses a new mechanism for the avalanches and plastic flow in the crystal plasticity.

**Methods**

*Sample synthesis and characterization.* The HKUST-1 crystals were synthesized through a hydrothermal method with the following procedure.[46] First, the cupric nitrate hemipentahydrate (2.327 g) was dissolved into the deionized water (25 mL), while the benzene-1,3,5-tricarboxylic acid (1.414 g) was added to the solvent (50 mL) containing the ethanol and deionized water equally. The resultant solution mixture was then transferred into a Teflon-lined stainless steel autoclave (250 mL), which was kept at 150 °C for 15 h. Last, after the autoclave was cooled down to room temperature, the yielded blue crystals were isolated by filtration. The formation of crystalline HKUST-1 was confirmed by the powder X-ray diffraction patterns at a scan speed of 5 °min$^{-1}$ and in the range $2\theta$ = 5-50 ° using the Cu-K$_\alpha$ radiation source (D-MAX 2500, Rigaku). The size and the topology of the synthesized HKUST-1 crystals were determined by the optical microscope (DM2700 M, Leica) and confocal microscope (S neox, Sensofar).

*In situ compression tests*. The HKUST-1 crystals were dispersed into ethanol solutions with the aid of the ultrasonic vibration, which were subsequently dripped on a freshly cleaved atomically flat silicon substrate. After the solvent was completely evaporated, HKUST-1 crystals were left and adhered to the substrate, which were activated in a vacuum oven at 100 °C before testing. The *in situ* compression tests were performed by the FemtoTools *in situ* SEM nanoindenter (FT-NMT04) with a diamond flat punch tip having a diameter of 10 μm, which was visualized by the SEM (Clara, Tescan) operated at the accelerating voltage of 5 kV. After the HKUST-1 crystal and the indenter tip were aligned with each other well, the indenter tip was moved towards the crystals with a rate of ~10 nm/s by using the constant loading rate indentation mode.

*Instrumented nanoindentation tests.* Nanoindentation experiments were also performed by the FemtoTools *in situ* SEM nanoindenter inside the Clara SEM. But a diamond Berkovich indenter with a tip radius of ~20 nm was used here. After an HKUST-1 crystal was located exactly below the indenter tip, it was indented at the constant strain rate of $1 \times 10^{-2}$ $s^{-1}$ with the thermal drift maintained below ±0.05 nm/s. Here, to eliminate the impact of the electron beam irradiation, the electron beam was switched off during the indentation process. After achieving the load-depth curves from the nanoindentation experiments, the Young's modulus of HKUST-1 crystals was calculated through the Oliver-Pharr method.[47]

*MD simulations.* A {111}-faceted HKUST-1 particle with the octahedral morphology was considered in current MD simulations. Here, the particle had an edge length of ~21 nm, which contained 191,568 atoms. In MD simulations, the interactions between atoms in HKUST-1 were described by the *ab initio* derived force field MOF-FF,[48] which has the capacity to simulate the pressure-induced phase transformations of MOFs.[49] To implement the compression simulations, the HKUST-1 particle was placed between two rigid planar indenters, which were, respectively, parallel to the top and bottom {111} facets of particles. A repulsive potential was employed to describe the frictionless compression between the particle and rigid planar indenters.[50] Prior to compression, the HKUST-1 particle was firstly relaxed using the conjugate gradient method to reach the stable configuration with a minimal energy, which was subsequently equilibrated at the temperature of 300 K for about 100 ps. In the compression process, the planar indenters were gradually moved towards each other with a constant speed of 0.01 Å/ps. All MD simulations conducted here were implemented by the open-source code LAMMPS[51] with the NVT ensemble. A velocity-Verlet integration with a time step of 1.0 fs was utilized in MD simulations, while the room temperature was controlled by using the Nosé-Hoover thermostat.

*FEM simulations.* The commercial software ABAQUS/Standard was used to conduct FEM simulations. Here, the edge length of the octahedral particle was set as 10 μm, close to that considered in the *in situ* compression tests, while the flat indenter tip modelled as a cuboid with the dimension of 16 μm × 16 μm × 10 μm was built above the particle (Figure S11). Both the HKUST-1 particle and indenter tip were described by the eight-node hexahedral solid elements (C3D8R). The number of elements is around 30,000 for the HKUST-1 particle. To implement the compression simulations, the displacement of the bottom surface of particles was constrained, while the rigid indenter was moved towards the particle with a rate of 10 nm/s, which is similar to that employed in experiments. After the indenter and particle were in contact with each other, the load and displacement were recorded. Here, the indenter was described as a discrete rigid body by setting an extremely large Young's modulus. Meanwhile, the material properties used in FEM simulations of the HKUST-1 particle were extracted by fitting the load-displacement curve obtained from simulations to that measured in experiments. In doing this, the initial stage of the load-displacement curve was used to approximate the elastic properties, which were assumed to be isotropic. Once the elastic properties were determined, the isotropic plastic properties were added to the model. Here, the elastic-perfectly plastic model was employed for the plasticity of HKUST-1 particles.

**Acknowledgments**

This work was supported by the Guangdong Basic and Applied Basic Research Foundation (Grant No. 2022A1515010631) and Shenzhen Science and Technology Program (Grant No. GXWD20220811164345003).

**Author contributions**

J.Z. conceived the idea and was involved in experiments and simulations. J.K. conducted the experiments. B.W. was involved in the sample preparation and conducted FEM simulations. X.M. conducted MD simulations. All authors were involved in the discussion of the results and the writing of the manuscript.

**Competing interests**

The authors declare no competing interests.

**Figures**

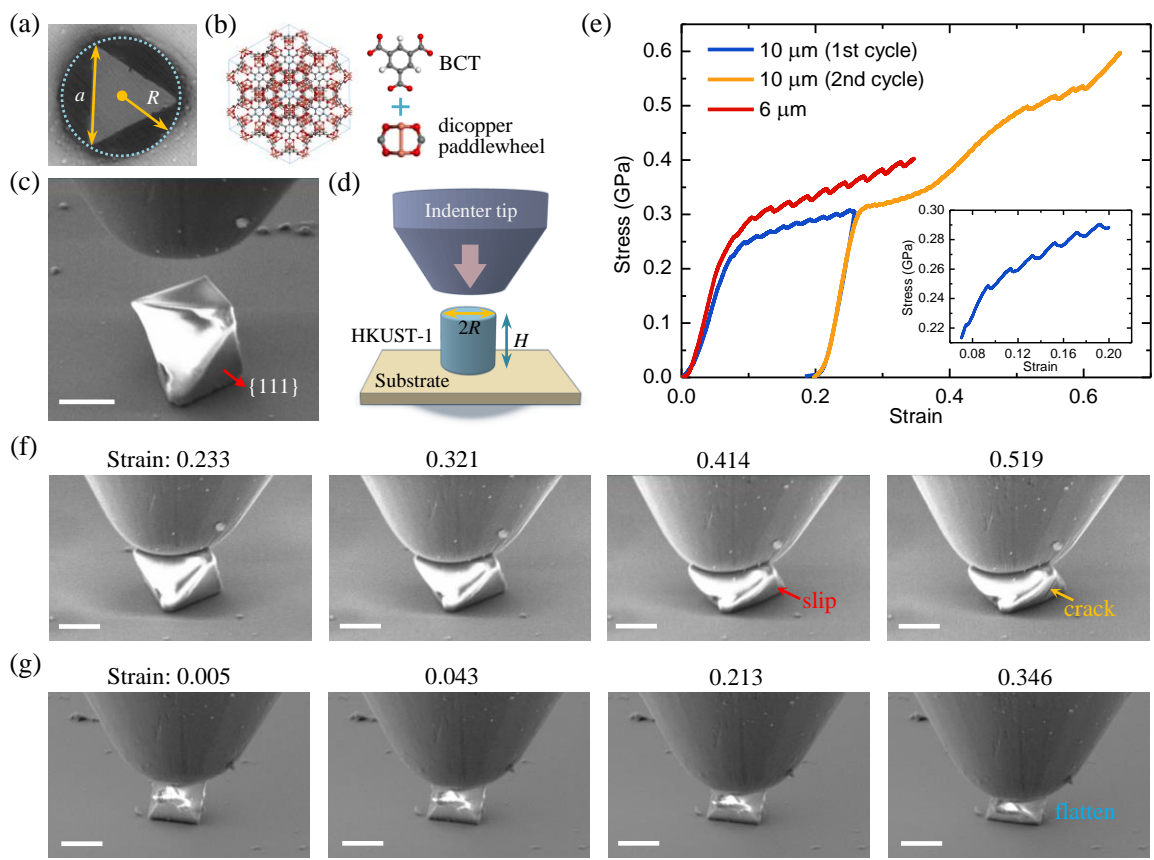

**Figure 1.** *In situ* compression tests of the individual HKUST-1 crystals. (a) Surface profile of the HKUST-1 crystal. (b) Atomic model of HKUST-1 along the {111} orientation together with its building units. (c) SEM image of an individual HKUST-1 crystal under the indenter tip, which was aligned for compression tests. (d) Schematic of the geometry simplification of the HKUST-1 crystal from octahedron to column. (e) Stress versus strain plots for the compression of crystals with different edge lengths. The inset highlights the serrated flow in the plasticity of the crystal with an edge length of 10 μm. (f) SEM images of the crystal with an edge length of 10 μm at various strains during the second round of compression. (g) SEM images of the crystal with an edge length of 6 μm at various strains. All scale bars are 5 μm.

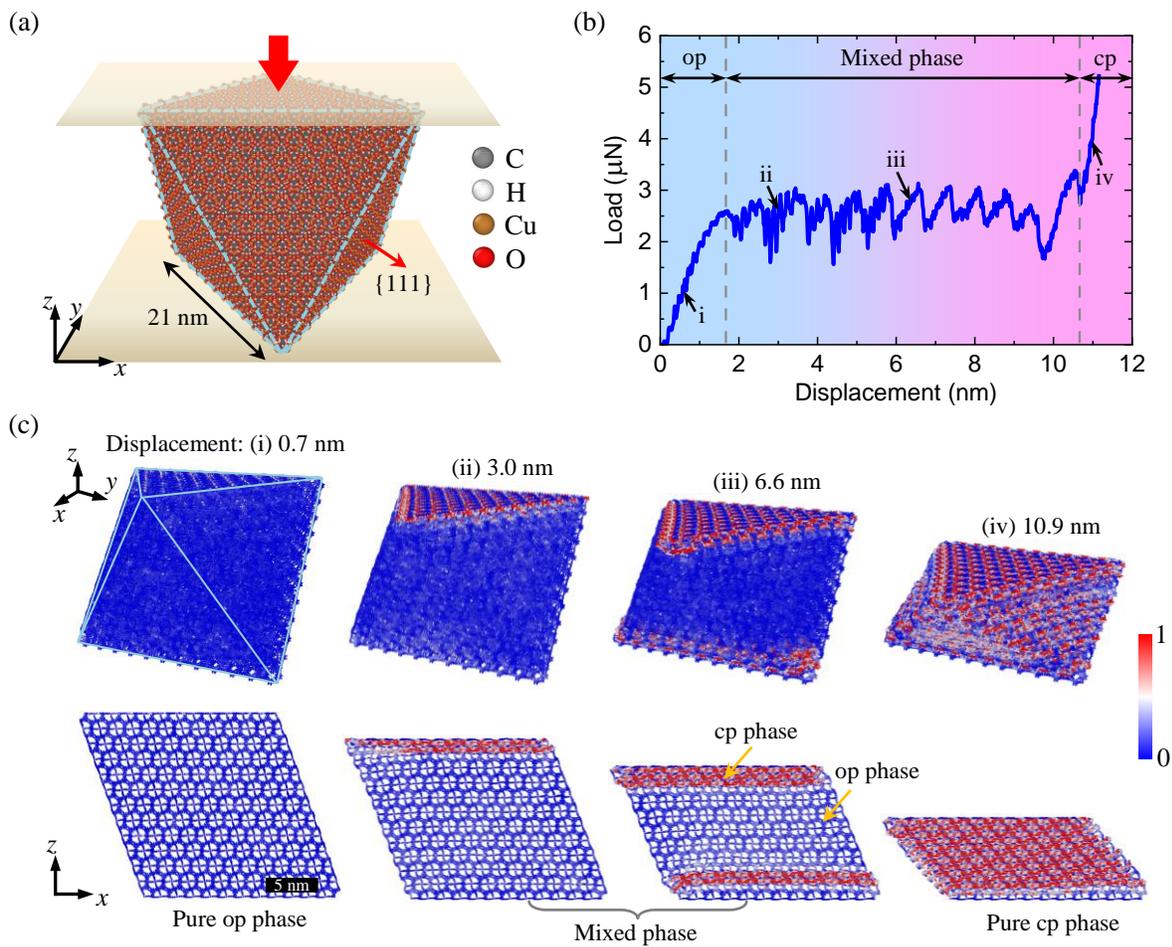

**Figure 2.** MD simulations for the compression of an individual HKUST-1 crystal. (a) Simulation setup for the compression of an HKUST-1 crystal. (b) Load versus displacement curve of the crystals during the compression process. (c) Perspective (top panel) and cross-sectional (bottom panel) views of the atomic strain in the HKUST-1 crystal compressed at different displacements.

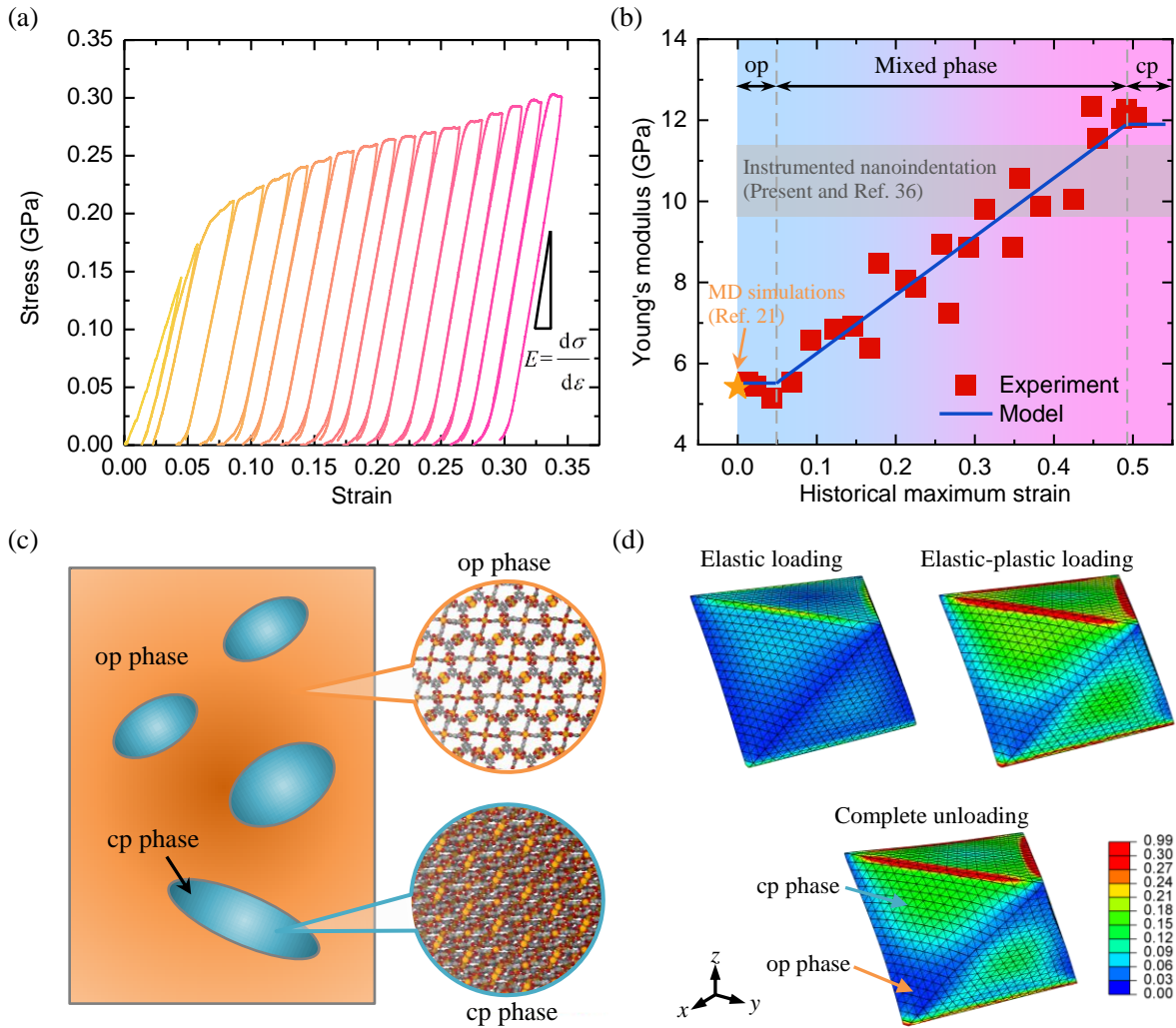

**Figure 3.** Elastic property of HKUST-1 crystals. (a) Repeated loading-unloading processes of the HKUST-1 crystal. The slope of each stress-strain ($\sigma$-$\varepsilon$) curve in the unloading process equals to the corresponding Young's modulus $E$. (b) The Young's modulus of HKUST-1 crystals previously compressed with various historical maximum strains. Here, results extracted from previous MD calculations[21] and measured from previous[36] and present IITs are also shown for the sake of comparison. (c) A composite model comprised of the op and cp phases, equivalently representing the HKUST-1 model with phase coexistence. (d) Strain distribution in the HKUST-1 crystal under the elastic and elastic-plastic loading and after the complete unloading.

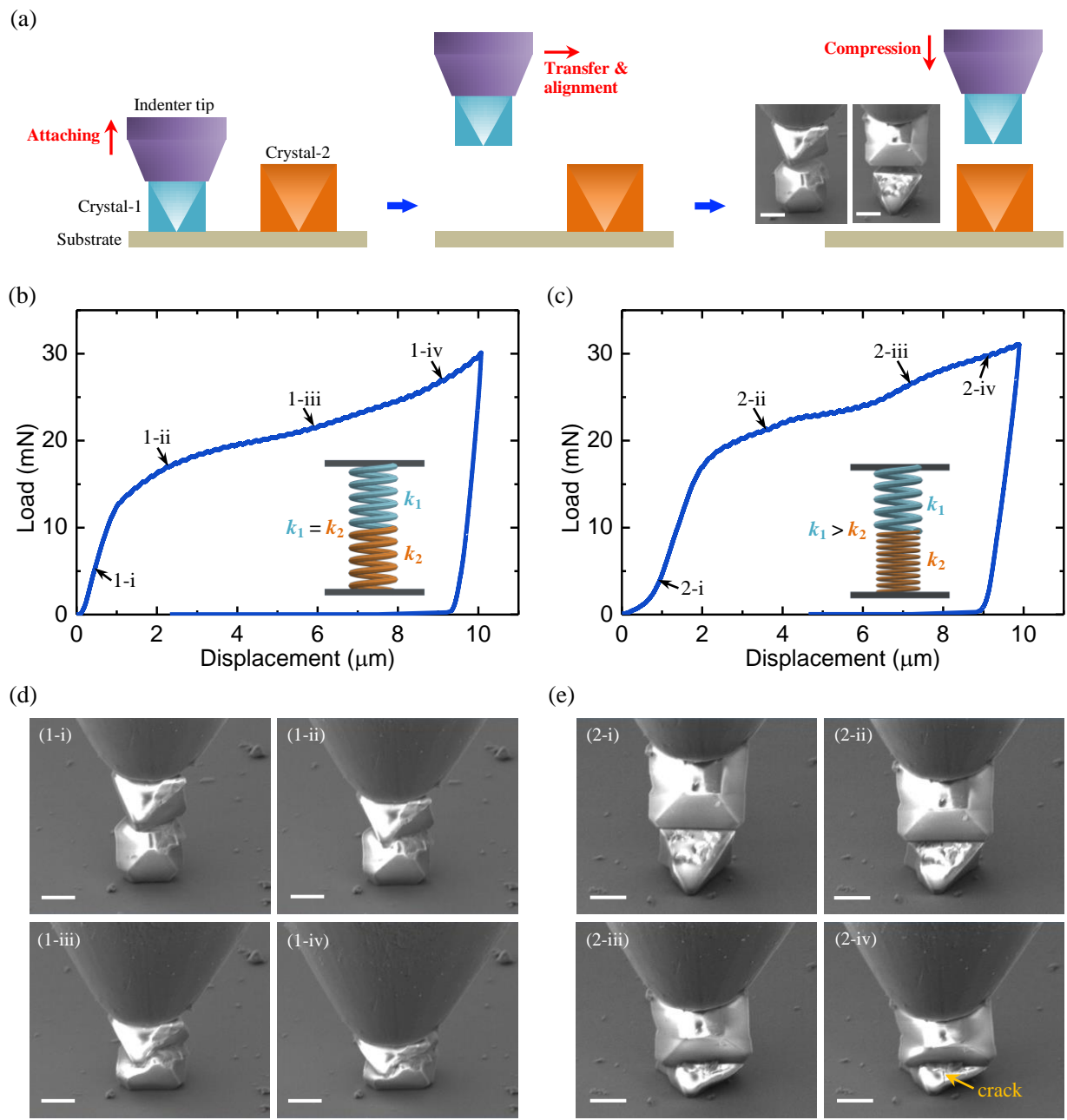

**Figure 4.** *In situ* compression tests of contacting HKUST-1 crystals. (a) Schematic of the particle transfer and alignment procedure for the compression tests of contacting HKUST-1 crystals. The insets illustrate the SEM images of the contacting crystals with comparable and different sizes considered in compression tests. (b) Load versus displacement curve for the contacting crystals with comparable sizes and (c) the corresponding result of their counterparts with different sizes.

(d, e) SEM images of the contacting HKUST-1 crystals compressed at various displacements as noted in (b, c). All scale bars are 5 μm.